# Awareness of crash risk improves Kelly strategies in simulated financial time series

Jan-Christian Gerlach[1], Jerome Kreuser[2], and Didier Sornette[1,3]

[1] Department of Management, Technology and Economics, ETH Zurich, Zurich, Switzerland

[2] RisKontroller Global LLC, Arlington, Virginia, USA

[3] Institute of Risk Analysis, Prediction and Management (Risks-X), Academy for Advanced Interdisciplinary Studies, Southern University of Science and Technology (SUSTech), Shenzhen, 518055, China

23 March 2020

## Abstract

We simulate a simplified version of the price process including bubbles and crashes proposed in Kreuser and Sornette (2018). The price process is defined as a geometric random walk combined with jumps modelled by separate, discrete distributions associated with positive (and negative) bubbles. The key ingredient of the model is to assume that the sizes of the jumps are proportional to the bubble size. Thus, the jumps tend to efficiently bring back excess bubble prices close to a "normal" or fundamental value ("efficient crashes"). This is different from existing processes studied that assume jumps that are independent of the mispricing. The present model is simplified compared to Kreuser and Sornette (2018) in that we ignore the possibility of a change of the probability of a crash as the price accelerates above the normal price. We study the behaviour of investment strategies that maximize the expected log of wealth (Kelly criterion) for the risky asset and a risk-free asset. We show that the method behaves similarly to Kelly on Geometric Brownian Motion in that it outperforms other methods in the long-term and it beats classical Kelly. As a primary source of outperformance, we determine knowledge about the presence of crashes, but interestingly find that knowledge of only the size, and not the time of occurrence, already provides a significant and robust edge. We then perform an error analysis to show that the method is robust with respect to variations in the parameters. The method is most sensitive to errors in the expected return.

# 1. Introduction

The optimal investment problem is concerned with allocating a portfolio of assets today with the expectation of obtaining "optimal results" in the future. A favourable strategy is to maximize expected portfolio growth. This objective is also known as the Method of Kelly (Kelly Criterion). Kelly (1956) showed that the strategy maximizes long-run wealth asymptotically on repeated investments over time. Much has been written and researched on the method ever since. The reader is referred to the compendium of papers in MacLean et. al. (2010a). Furthermore, the promising strategy has not only been extensively studied, but also been successfully applied by many (see papers in L. C. MacLean, Edward O. Thorp, and W.T. Ziemba (2010) and Ziemba (2016) for an excellent discussion and review of Kelly).

Much theory on and applications of the Kelly Capital Growth Investment Strategy model asset prices as Geometric Brownian Motion (GBM). We know this assumption does not generally hold for real financial time series. In a Gaussian view of the world, the occurrence of extreme events is strongly underestimated. Ait-Sahalia and Matthys (2018) demonstrate that ignoring this uncertainty leads to significant loss of wealth for the investor. Therefore, strategies that assume normal returns fail at accurately pricing this risk and instead result in inappropriately risky and overly leveraged positions. Hence, oftentimes it is considered more reasonable to model price processes by fat-tailed distributions such as the stretched-exponential and other distributions (see Malevergne et al., (2003; 2004; 2005)). These models allow rare jumps to occur more often and thus more accurately fit real financial returns.

An often-followed approach is to model asset prices as Geometric Brownian Motion, augmented by a jump process that accounts for the occurrence of extreme events. Commonly, jumps are modelled as a class of independent and identically distributed Poisson random measures. Much work has been published on this type of asset price process (Sévi and Baena (2011); Ait-Sahalia and Jacod (2009); Bass (2004); Cont and Tankov (2004); Papapantoleon (2008), to name a few).

But including jumps may not be sufficient to account for the arguably most important investment risk, namely a crash whose amplitude is beyond the fat-tailed distribution and can be considered as an outlier (Johansen and Sornette (2001/2002)). Typically, the most severe events that cause investors to incur large losses in their portfolios are indeed price crashes ending financial bubbles. Nevertheless, bubbles are often characterized by remarkable growth and thus potentially offer rewarding gains in advance to the crash. Investors thus find themselves playing the difficult game of deciding between the temptation to exploit bubble growth and the risky exposure to crashes.

Naturally, the risk of financial bubbles and crashes should directly impact the optimal investment strategy of investors. However, strategies that only assume normal returns such as the classical Kelly approach, or even fat-tailed return distributions, fail to incorporate this type of risk and therefore blindly run into crashes. Then, it becomes simply a "question of luck" for an investor whether the gains made during the advancement of the bubble are sufficiently high to compensate the losses encountered due to the crash, where luck here signifies the "random" time during the bubble growth phase at which the investor entered the market and the stochastic amplitude of the crash.

Instead of exposing ourselves to this fate, we rather seek for strategies that effectively exploit bubble growth while mitigating (or even shorting) crashes and at the same time maintain the positive property of the classical Kelly strategy of maximizing the expected log of wealth. Such an augmented Kelly strategy which satisfies this need has been derived for and presented together with the "Efficient Crashes Model" that was proposed by Kreuser and Sornette (2018).

The asset price process underlying the Efficient Crashes Model is defined as a combination of GBM and jump process, with a price-dependent jump hazard rate. The price of the asset nonlinearly oscillates about a slow-moving normal price process in a (highly nonlinear) mean-reverting manner (see the normal price as defined by Merton (1971)). The mispricing between asset and normal price is the size of a forming positive or negative bubble. The bubble is ultimately ended by a single jump or a series of jumps originating from the jump process. The jump size is relative to the size of the mispricing. In this sense, the crashes (endings of positive bubbles) or up-jumps (endings of negative bubbles) are efficient, as they are directly proportional to the size of the bubble and always drive back the price towards the stable normal price. This formulation allows to quantify the risk of a crash firstly by its magnitude and secondly by the estimated probability of a jump to occur in time.

In Kreuser and Sornette (2018), an important ingredient consists in a dynamic model for the probability of a jump event as a function of the mispricing (bubble size). Our approach here is different. We study a reduced constant-parameter version of the model, in which we omit the dynamic probability of jumps and replace it by a constant value. Thus, we assume that jumps are Poissonian in their temporal occurrence. Then, although the crash time is not itself predictable anymore, the crash size still is, as it is directly related to the bubble mispricing. So, an asset with a higher mispricing can be interpreted to be riskier, as it will likely experience larger crashes.

Thus, the model considered in this work covers "half the ingredients" of Kreuser and Sornette (2018) by keeping the efficient crash effect (i.e. the dependence of the crash size on the mispricing or size of the bubble), but removing any dependence of the jump probability on other model parameters. What is interesting is that, notwithstanding the absence of predictability of occurrence times of the crashes, the predictability of only their sizes already provides a competitive advantage over other strategies that do not incorporate this risk.

To demonstrate this, in the following, we carry out various simulation studies based on synthetic price series, in order to gain insights about the model dynamics, the impact of parameter estimation error and in particular the augmented Kelly procedure built around the core model, as this has most interesting implications for real-world financial applications. We do so by performing (i) a comparison of our generalized Kelly method to the classical Kelly formulation for GBM price processes, (ii) a parameter sensitivity analysis and ultimately (iii) an error robustness analysis. These are successively presented in the following sections.

In the next section, we review and summarize the method developed in Kreuser and Sornette (2018). In Section 3, we discuss how the simulation sample paths are generated. In Section 4, we compare our strategy with other strategies including a 60/40 portfolio and classical Kelly. In Section 5, we discuss the sensitivity of the model output with respect to the parameters and, in Section 6, we do an error analysis defining the degree of robustness that we achieve. Section 7 concludes.

2. **Reduced Efficient Crashes Model**

We define the following set of variables:

$\Delta t$ = discrete time interval $[t, t+1]$.

$p_t$ = price of the risky asset at time t.

$\bar{r}_t$ = expected return of the risky asset on $\Delta t$ when there is no crash or rally.

$\sigma$ = standard deviation on $\Delta t$ of the geometric random walk price process.

$\varepsilon_t$ = sample from a standard normal distribution at time t.

$r_D$ = discount rate of the asset price on $\Delta t$.

$r_N$ = growth rate of the "normal price" on $\Delta t$.

$r_f$ = risk-free rate on $\Delta t$.

$p_0$ = starting price of the risky asset.

$N_t = p_0 \, exp(r_N t)$ : this defines the normal price process.

$\rho_t$ = probability that there is a correction (crash or rally) at time t.

$\kappa_i \in (-\infty, \infty)$ = the size of the i[th] corrective jump relative to the distance to the normal price.

$\eta_i$ = probability that, when there is a correction, it is of size $\kappa_i$.

$\bar{K} \equiv \sum_{i=1}^{n} \eta_i \kappa_i$ is the expected corrective crash size relative to the distance to the normal price.

$q_t = \frac{N_t}{p_t}$ is the relative (inverted) mispricing of the risky asset.

We follow Kreuser and Sornette (2018) and use the discrete stochastic price process:

$$p_{t+1} = p_t \exp(\bar{a}_t + \sigma \varepsilon_t) \quad \text{with} \quad p_0 > 0$$

and

$$\bar{a}_t = \begin{cases} \bar{r}_t & \text{with probability } 1-\rho_t \quad \text{with } 0 \le \rho_t < 1 \\ \kappa_i \ln(q_t) + r_D & \text{with probability } \rho_t \eta_i \quad i=1,2,\ldots,n \\ & \kappa_i \in \Omega \equiv \{\kappa_i | -\infty < \kappa_i < \infty, \ i=1,2,\ldots,n\} \end{cases} \quad (1)$$

with

$$q_t = \frac{N_t}{p_t} \quad \text{and} \quad \sum_{i=1}^{n} \eta_i = 1 \quad 0 < \eta_i < 1 \quad \text{and} \quad \bar{K} = \sum_{i=1}^{n} \eta_i \kappa_i$$

$$N_t = p_0 \exp(r_N t)$$

The normal price process $N_t$ follows Merton's (1971) "asymptotic normal price-level" that assumes there is a price function $N_t$ with

$$\lim_{t \to \infty} E_T \left[\frac{N_t}{p_t}\right] = 1 \quad \forall \ 0 \le t \le T < \infty \quad (2)$$

We do not assume that our normal price is a fundamental price. Therefore, our normal price does not require an additional, external pricing model to calculate it, but is simply an output of the bubble calibration. Majewski et al. (2018) extend a Chiarella model with a fundamental price as an output of the calibration that also does not require an external pricing model. Yan et al. (2010) infer a fundamental value from the bubble calibration.

We assume in addition that the expected return is determined in accordance with the Rational Expectations or efficient crashes condition

$$E_t \left[\ln \frac{p_{t+1}}{p_t}\right] = r_D \quad \forall t \quad (3)$$

In the case where the crash probability is constant over time $E_{t-1}[\rho_t] = E[\rho_t] = \bar{\rho}$, we obtain

$$\begin{aligned} E_t \left[\ln\left(\frac{p_{t+1}}{p_t}\right)\right] &= (1-\bar{\rho})\bar{r}_t + \bar{\rho}\left(\sum_{i=1}^{n} \eta_i \kappa_i\right) \ln(q_t) + \bar{\rho} r_D \\ &= (1-\bar{\rho})\bar{r}_t + \bar{\rho}\bar{K} \ln\left(\frac{N_t}{p_t}\right) + \bar{\rho} r_D \\ &= r_D \end{aligned} \quad (4)$$

where $\bar{K}$ is the expected crash factor already defined in (1). With the RE equation, the value $\bar{r}_t$ of the expected return of the risky asset is:

$$\bar{r}_t = r_D - \frac{\bar{\rho}\bar{K} \ln q_t}{1-\bar{\rho}} \quad (5)$$

If there is never a crash ($\bar{\rho} = 0$), then the expected return of the risk asset is always $r_D$.

For this study, we will use the constant probability assumption throughout the rest of the paper. In contrast, in Kreuser and Sornette (2018), we allow for the jump probability $\rho_t$ to increase with price, leading even great price acceleration. This is possible because returns are easier to estimate as they increase in size relative to their volatility (See Merton, 1980).

Equation (5) is the key to the Efficient Crashes method and the parameter relationships, both in Kreuser and Sornette (2018) and in the present, simplified version. As the mispricing $\frac{1}{q_t} = \frac{p_t}{N_t}$ increases, so does $\bar{r}_t$, as seen from Equation (5), which then feedbacks on the price process (1), leading to a faster-than-exponential growth

$$p_{t+1} = \frac{1}{N_t}\left(\frac{p_t}{N_t}\right)^{1+\frac{\bar{\rho}\bar{K}}{1-\bar{\rho}}} e^{\sigma\varepsilon_t} \tag{6}$$

as seen from the value of the exponent $1 + \frac{\bar{\rho}\bar{K}}{1-\bar{\rho}}$ which is larger than 1 for non-zero $\bar{\rho}$ and $\bar{K} > 0$. In the present, reduced model, this feedback loop is the only mechanism leading to transient, super-exponential bubbles. Kreuser and Sornette (2018) added an accelerating jump probability $\rho_t$ as a function of price, while here we take $\rho_t$ to be constant and equal to $\bar{\rho}$, as already mentioned.

Summarizing, our approach captures several stylized properties of asset prices including:

- Heavy tails (Johansen and Sornette (2001); Lux and Sornette (2002));
- Asymmetry in crashes versus rallies (Cont (2000));
- Momentum plus mean-reversion (Yan et al. (2012), Majewski et al. (2018));
- Super-exponential growth (Sornette (2003); Sornette and Cauwels (2015); Leiss et al (2015));
- A normal price in the sense of Merton (1971).

Next, we apply the method of Kelly to the price process in (1) for a two-component investment setting, i.e. a situation where we can choose to allocate our capital between a risky and a risk-free asset. Let $\lambda_t$ be the fraction of wealth $W_t$ allocated to the risky asset in time $t$ and $1 - \lambda_t$ the allocation to the risk-free asset with return $r_f$. Then

$$W_{t+1} = (\lambda_t e^{\bar{a}_t + \sigma\varepsilon_t} + (1-\lambda_t)e^{r_f})W_t \tag{7}$$

where $\bar{a}_t$ has been defined in (1). According to the Kelly criterion, we wish to determine

$$L(\lambda_t^*) := \max_{\lambda_t} E_t\left[\ln\frac{W_{t+1}}{W_t}\right] \tag{8}$$

where $E_t$ is the expectation conditional on information up to time $t$. Numerically solving for this maximized objective function in each time step yields a series of optimal investment fractions with respect to the Kelly criterion. In Kreuser and Sornette (2018), several desirable properties of the objective function as well as an approximate solution for the optimal investment fraction (see Appendix A in this paper) are presented.

## 3. Synthetic Data

In the following, we describe the synthetic data generation procedure. A simulation of classical Kelly based methods, applied to pure Geometric Brownian Motion only, was presented by Wesselhoft (2016) and useful insights such as closed-form solutions were derived. Here, we study the performance of our augmented Kelly strategy on datasets generated by the simplified version of the Efficient Crashes Model (ECM) introduced in the preceding section. These time series serve as surrogate datasets for performance evaluation of the Efficient Crashes Optimizer Kelly portfolio strategy, as well as other standard portfolio allocation strategies that serve as benchmarks.

Note that, throughout the paper, we deliberately and exclusively study synthetic datasets. The advantage of synthetic data simulations is that there is no model error. We can thus directly investigate the impact of estimation error arising from the involved parameter estimation procedures. The success of the method has already been demonstrated for numerous real-world datasets in the original paper of 2018. Therefore, we refer any reader who is interested in an application of the model to actual financial time series to the original paper.

Together with the set of initial conditions $p_0 = p_{N,0} = 1$, we use the following, straightforward algorithm to obtain a dataset of length T.

**Generating Sample Paths**: Let $r_D, r_N, \sigma, \sigma_\kappa, \rho, \bar{K}$ be given.

For $t$ in 1,2,…,T:

1) Compute $q_t = \frac{N_t}{p_t}$
2) Draw $X_t \sim U(0,1)$
3) Draw $\varepsilon_t \sim N(0,1)$
4) If $X_t > \rho$:

$$p_{t+1} = p_t \exp\left(r_D - \frac{\rho \bar{K} \ln(q_t)}{1-\rho} + \sigma \varepsilon_t\right) \quad (9)$$

5) Else if $X_t \leq \rho$:
   a. Draw $\kappa_t \sim N(\bar{K}, \sigma_\kappa^2)$
   b. Let $p_{t+1} = p_t \exp(r_D + \kappa_t \ln(q_t) + \sigma \varepsilon_t)$ \quad (10)
6) Compute $N_{t+1} = N_t \exp(r_N)$

where $U$ and $N$ indicate standard uniform and normal distributions. All generated paths are initialized in a 'no-bubble' state because the initial price equals the initial normal price. Usually, we estimate the initial mispricing, as well, because the asset may already be in a bubble state at the beginning of the analysed timeframe.

The model comprises a variety of parameters that need to be specified in advance, namely the price process rates $r_N$ and $r_D$, the volatility of the Gaussian Error term $\sigma$, the probability of a jump $\rho$, the average corrective jump size $\overline{K}$ and the standard deviation of the jump distribution $\sigma_\kappa$.

As reference values for these parameters, we pick a set of base values:

$$\Phi = \left\{r_N = r_D = \frac{\ln(1.07)}{252}, \sigma = \frac{0.17}{\sqrt{252}}, \rho = 0.01, \overline{K} = 0.3, \sigma_\kappa = 0.2\right\} \quad (11)$$

We keep these "true values" constant during our simulations. The values for the rates and volatility, i.e. roughly 7% yearly growth at 17% annual volatility, are derived from underlying S&P 500 Index historical data using sample mean and volatility of the entire time series between 1971 and 2019[1]. Kreuser and Sornette (2018) point out that it is not unreasonable to assume that $r_N$ and $r_D$ are equivalent, at least in the long run. The base value for the constant jump probability is chosen such that every 100'th trading day we can expect a jump to occur. The average corrective crash size $\overline{K}$ is set to about one third of the mispricing. We do not generally expect the price of an asset in bubble state to perfectly correct back towards the fundamental value once a crash or up-jump appears. Thus, a value of $\overline{K} < 1$ is a reasonable choice to represent this behaviour in the data. The individual jump magnitudes $\kappa$ are realizations drawn from a normal distribution with mean $\overline{K}$ and standard deviation $\sigma_\kappa$. The standard deviation is here chosen such that about 70% of the drawn jumps are within $0.1 - 0.5$ times the size of the mispricing component[2] and therefore variation in the size of the jumps relative to the mispricing is allowed.

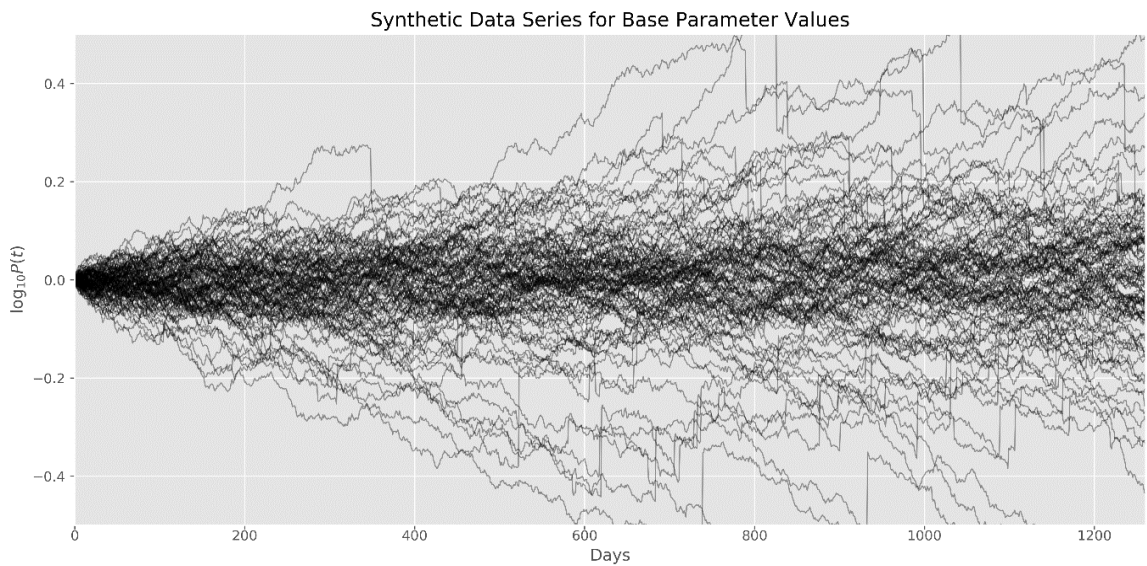

**Figure 1:** Simulation of 100 example trajectories at the base parameter values (11).

---

[1] It is not crucial that the values we pick exactly match the true values of the S&P500 dataset. They should be simply of an order that approximately represents real-world behaviour.
[2] There is no empirical evidence that these latter values apply, however for demonstrational purposes, we consider them as realistic values. Furthermore, as will be seen below, we also show the model performance over ranges of the true parameter values, such that the reader may comprehend how the model behaviour changes depending on these values.

Figure 1 shows 100 synthetic datasets of duration $T = 1250$ (i.e. five years of business day frequency data) that were simulated using the base values $\Phi$ (11). Furthermore, in the Appendix, in Figure 18 - Figure 20**Figure *3***, we explore variations in the data by changing single parameters while keeping the others fixed at their base values. In order to increase the comparability between the plots, the same random seed is chosen for all simulations, such that we can reproduce the results at any time. Note that the parameters $\rho$ and $\bar{K}$ of the jump process greatly influence the appearance of synthetic data. As we increase their values, we see a drastic departure from the pure GBM behaviour towards more erratic and 'bubbly' price trajectories. Overall, the presented plots demonstrate that the synthesis procedure generates data as expected, i.e. there are no abnormal or unexpected features in the data[3].

In the following section, we compare the Efficient Crashes Optimizer (ECO) Kelly strategy with several standard portfolio strategies on synthetic datasets generated by means of the algorithm above. We assume the true parameters to be known which corresponds to perfect estimation of the quantities. This will be relaxed later. The performance of all strategies will be compared over increasing windows of time of the underlying datasets. We will show that ECO Kelly outperforms all other strategies in the long run, given that the data originates from the efficient crash bubble model.

## 4. Comparison of Strategy Performances

We compare several strategies to the ECO portfolio:

- **Buy and Hold (B&H):** exactly replicates the returns on the risky asset.
- **Fixed Fraction (FF):** invests 60% of available capital into the risky asset and the remainder into the risk-free asset, also commonly known as 60/40 portfolio.
- **Classical Kelly (CK):** assumes the data originates from a pure GBM process, investing according to the corresponding closed-form solution for the optimal Kelly fraction (see below).

For all simulations, we set the risk-free rate $r_f = 0$. A portfolio can be simulated based on the series of investment decisions $\{\lambda\}_t$ taken at each time step. We defined the investment fraction $\lambda_t$ as the fraction of our capital allocated to the risky asset at time $t$ over the next interval $[t, t + \Delta t]$. Accordingly, in a two-component-setting comprising a single risky and a risk-free asset, $1 - \lambda_t$ is the fraction of money that remains invested in the risk-free asset. We assume that our portfolio is rebalanced at every time step and that the initial value of our wealth is $W_0 = 1$ for all simulations. We compute the evolution of wealth (portfolio value) over a given timeframe as

$$W_t = W_{t-1}\big(\lambda_t e^{r_t} + (1 - \lambda_t)e^{r_f})\big) \qquad \text{for } t = 1, \dots, T \qquad (12)$$

---

[3] Note in particular the symmetry between positive bubbles followed by crashes (negative jumps) and negative bubbles followed by rallies (positive jumps). Real empirical data does not have this symmetry, since positive bubbles are significantly more numerous than negative bubbles, except in foreign exchange markets by the obvious symmetry that, e.g. EUR in USD is the inverse of USD in EUR by definition. We keep here the symmetry for the sake of simplicity of the model formulation.

whereby $r_t$ is the actual log-return of the risky asset. For the B&H strategy, we have $\lambda_t = 1\ \forall t$ and for the FF strategy, we have $\lambda_t = 0.6\ \forall t$. In the case of CK, we use the common solution $\lambda_t = \frac{\mu - r_f}{\sigma^2}\ \forall t$. In contrast to these benchmark strategies, due to the dependence on the mispricing, the ECO strategy results in a variable investment fraction $\lambda_{ECO}$.

Depending on the values of the model parameters, the estimated values of $\lambda$ can become quite large (positive or negative). If we restrict $\lambda \in [0,1]$, the Kelly strategy never risks ruin even with jumps. A value outside these bounds essentially means that we introduce leverage or shorting into our strategy, thus it becomes possible for the strategy to result in a negative wealth or bankruptcy. The more unconstrained $\lambda$ is, the more likely to outperform but the more volatile the strategy can be. In order to limit this risk, we compute both the ECO and the CK strategies for a constrained leverage situation where we limit $\lambda \in [-1,2]$ (i.e. we truncate our investment fraction such that we can go maximally 200% long or 100% short). For comparison, we also compute the completely unconstrained case where $\lambda \in [-\infty, \infty]$. The cost associated with rebalancing and leverage is neglected. Summarizing, in total, we obtain six strategies, one bounded and one unbounded strategy for ECO and CK, plus the B&H and Fixed Fraction (FF) strategies.

Figure 2 shows how the various strategies perform on sample price trajectories. The synthetic time series were generated using the base parameter values (11). The top and third-row panels of Figure 2 show the evolution of the price (equivalent to the B&H strategy) and wealth trajectories while the second- and last-row panels show the series $\{\lambda\}_t$ for each strategy. For the chosen base parameter values, the unbounded and bounded CK strategies result in the same investment fraction, because the expected return is constant and equal to the return on the normal price and $\lambda$ is in the range [-1,2]. Thus, in the plots, the bounded and unbounded strategies coincide (for the given choice of base parameter values). This is not the case in the ECO portfolio. Both ECO and CK make use of leverage and tend to outperform the B&H strategy during bubble phases. Due to the fixed investment fraction, FF & CK naively run into crashes, while for ECO, we see a tendency to decrease the investment fraction to lower absolute values during bubbles. During bubble phases, the ECO therefore trades at a lower leverage than CK which makes it less vulnerable to crashes. Nevertheless, we do see strong drawdowns in the ECO portfolio as well. The ECO portfolio does not perfectly circumvent crashes since the crash occurrences are purely Poissonian (constant probability per unit time) and thus not predictable, in contrast to the model studied in Kreuser and Sornette (2018).

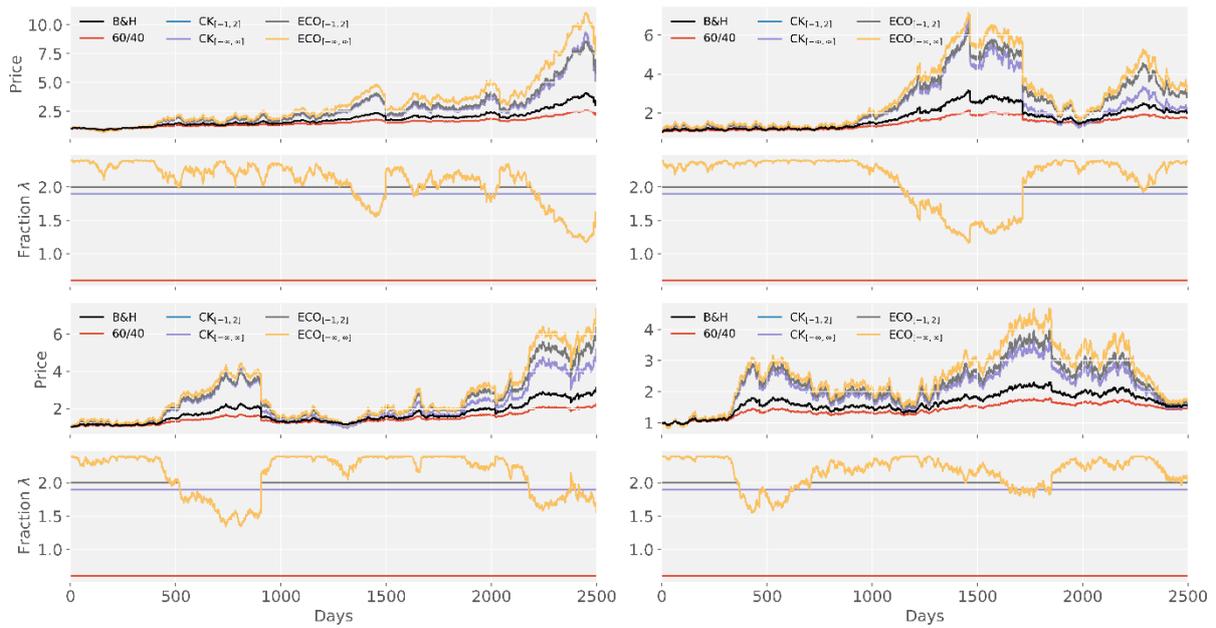

**Figure 2:** Four examples of portfolio evolution based on the synthetic datasets. The upper panels show the price / value of the portfolios that are constructed based on the different strategies. The lower panels show the associated investment fraction series for the ECO strategy, based on which the portfolios are computed. The black line (buy-and-hold-strategy) represents the evolution of the price of the underlying risky asset.

Next, we present another view of the simulation results for the ECO and CK portfolios. In separate simulations, we fix the synthetic data window duration T to 2 years (500 points) and 10 years (2,500 points) and simulate $m = 10,000$ time series for each window duration. We compute the log-performance of the B&H strategy, which is simply the terminal log-price $\ln P_T$ of the current price trajectory, given the initial price is $P_0 = 1$.

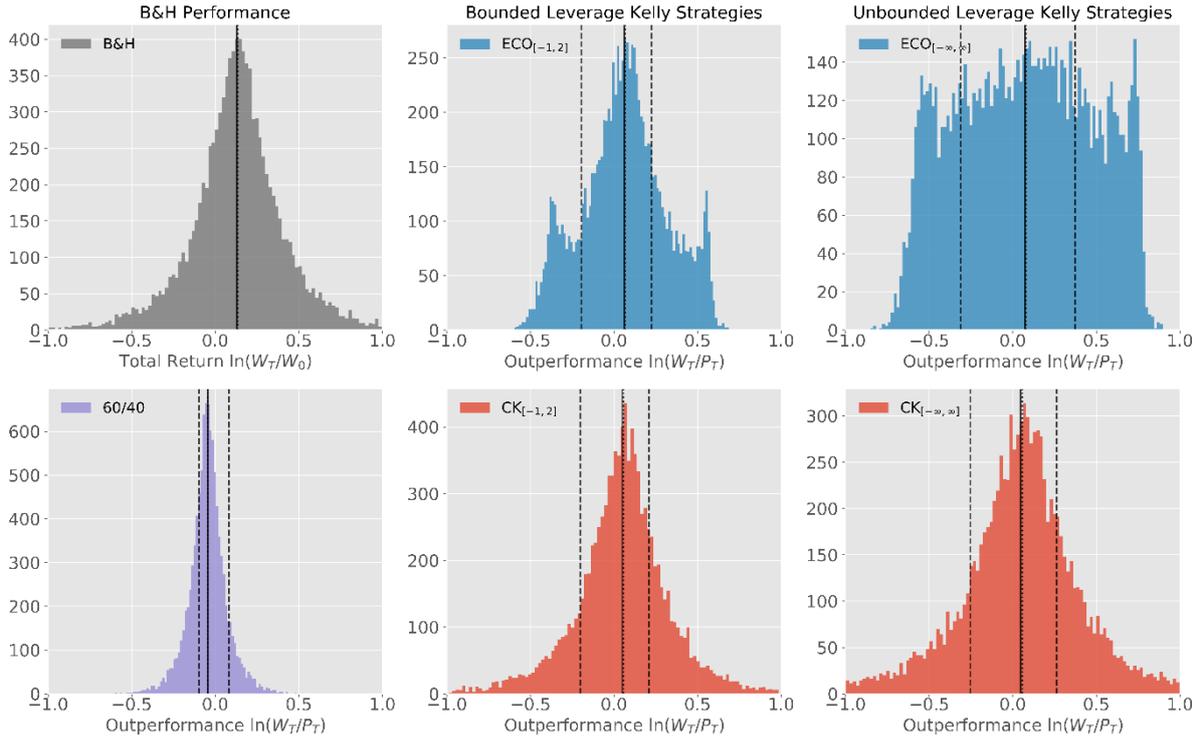

**Figure 3:** Outperformance histograms for the $T = 2$ years run of $m = 10{,}000$ simulations. The black vertical lines in each window indicate the mean (solid), the conditional negative / positive mean (left / right dashed), as well as the median (dotted).

|  | B&H | 60/40 | $CK_{[-1,2]}$ | $CK_{[-\infty,\infty]}$ | $ECO_{[-1,2]}$ | $ECO_{[-\infty,\infty]}$ |
|---|---|---|---|---|---|---|
| $Pr(W_T > P_T)$ [%] | - | 30.020 | 61.058 | 58.307 | 61.270 | 56.160 |
| Average Odds | - | 0.833 | 1.036 | 1.027 | 1.154 | 1.199 |
| Outperformance | - | -0.044 | 0.049 | 0.046 | 0.062 | 0.073 |
| $Pr(W_T = 0)$ [%] | 0 | 0.000 | 0.020 | 0.150 | 0.000 | 0.000 |
| Fraction of Uptime [%] | - | 27.545 | 64.671 | 62.475 | 65.070 | 59.980 |
| Sharpe Ratio (ann.) | 0.357 | 0.398 | 0.250 | 0.213 | 0.270 | 0.220 |
| CALMAR (mon.) | 0.038 | 0.043 | 0.026 | 0.022 | 0.028 | 0.024 |
| SDSR Sharpe (ann.) | 0.37 | 0.418 | 0.253 | 0.214 | 0.272 | 0.235 |
| CAGR [%/y] | 7.048 | 4.632 | 10.322 | 10.316 | 10.610 | 11.424 |

**Table 1:** Performance Metrics for the $T = 2$ years run

We compute the log-outperformance of 60/40, CK and ECO over the B&H strategy for each simulation:

$$O_i[T] = \left(\ln \frac{W_T}{P_T}\right)_i \quad \text{for} \quad i \in [1, \ldots, m] \tag{13}$$

where $i$ refers to the i'th simulation.

For each strategy, we additionally compute the following metrics:

(i) **Probability of Outperformance:** this is the probability $P_T$ that the terminal value of a strategy, $W_T$, is larger than the terminal value of the B&H strategy. It is the share of the histogram of terminal wealth that is positive.

(ii) **Average Odds:** this is the average return of outperforming strategies ($O_i > 0$) divided by the average return of underperforming strategies ($O_i < 0$). In the histograms, these averages are indicated by the right, respectively, left lines.

(iii) **Mean Outperformance:** this is the mean value of the outperformance at terminal time $t = T$ over all price series.

(iv) **Probability of Default:** this is the probability that the terminal value of a strategy, $W_T$, is zero. The probability of outperformance (i), the average odds (ii) and the probability of default are the classical measures that are considered for performance evaluation of Kelly strategies.

(v) **Fraction of Uptime:** this is the fraction of time that a strategy time series is located above the synthetic price time series of the risky asset. It is individually computed for each of the m simulations as $P(W_i > P_i)$ and then averaged over all simulations (this is also done for all metrics listed below). This metric can be seen as a more robust extension of metric (i), as it provides information about the performance throughout the whole simulated timeframe, rather than just the terminal time. It is also more robust with respect to variation of the start and end time of the timeframe[4].

(vi) **Sharpe Ratio:** the annualized Sharpe Ratio, as estimated from the log-returns of a given strategy time series by dividing their sample mean by the sample variance ($r_f = 0$). Essentially, by tuning our base parameters $\Phi$ (11), we can produce any desired value of SR. Thus, we advise not to interpret the absolute values of the attained SR's, but rather regard them in relation to each other (same is valid for the metrics listed below).

(vii) **California Managed Accounts Reports (CALMAR) Ratio:** the monthly CALMAR ratio, as an alternative risk-adjusted return-measure. It is the maximum month-to-month drawdown that we encounter in our portfolio.

(viii) **Symmetric Downside Risk Sharpe Ratio (SDRSR):** following Ziemba (2003) in the SDRSR, only the contribution of negative returns to the volatility is considered, as there is little reason to penalize on the upside.

(ix) **Compound Annual Growth Rate (CAGR):** the geometric growth rate of our portfolio, computed on a per-year base.

---

[4] Note that we could also think of this metric as being the probability of outperformance and metric (i) the probability of terminal outperformance. However, in order to differentiate better the naming, we chose to refer to metric (v) as fraction of uptime.

Next, we compute the histograms of the log-outperformances for the six strategies. The outcomes are depicted in Figure 3 and Figure 4, as well as in Table 1 and Table 2. Comparing the plots, we see that the distributions widen for the larger window duration (note that the x-scale changes), because the resulting portfolio returns for simulations of duration $T = 2500$ are larger than the ones for $T = 500$. Furthermore, the left tail of the CK strategies is heavier than in case of the ECO strategies. Moreover, for the long-time case, the medians and means of the distributions are further apart, thus the distributions become more skewed. These initial results reveal that the ECO dominates the CK and other strategies in most aspects.

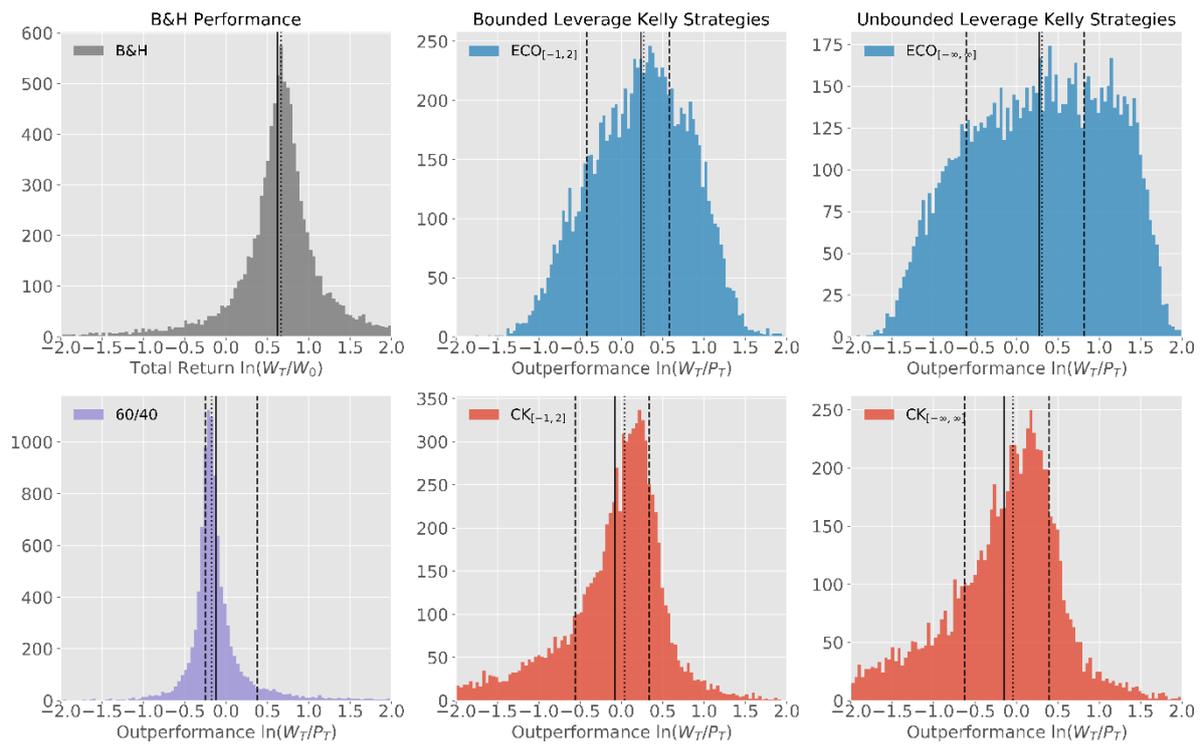

**Figure 4:** Outperformance histograms (same as figure 3) for the $T = 10$ years run of $m = 10,000$ simulations.

|  | B&H | 60/40 | CK$_{[-1,2]}$ | CK$_{[-\infty,\infty]}$ | ECO$_{[-1,2]}$ | ECO$_{[-\infty,\infty]}$ |
|---|---|---|---|---|---|---|
| Pr($W_T > P_T$) [%] | - | 20.012 | 53.454 | 46.909 | 65.369 | 61.660 |
| Average Odds | - | 1.530 | 0.605 | 0.632 | 1.396 | 1.351 |
| Outperformance | - | -0.122 | -0.079 | -0.147 | 0.237 | 0.273 |
| Pr($W_T = 0$) [%] | 0 | 0.000 | 10.620 | 14.840 | 1.250 | 1.250 |
| Fraction of Uptime [%] | - | 14.914 | 57.697 | 50.180 | 73.930 | 64.994 |
| Sharpe Ratio (ann.) | 0.257 | 0.324 | 0.121 | 0.081 | 0.221 | 0.189 |
| CALMAR (mon.) | 0.018 | 0.025 | 0.006 | 0.002 | 0.015 | 0.013 |
| SDSR Sharpe (ann.) | 0.25 | 0.330 | 0.113 | 0.076 | 0.197 | 0.174 |
| CAGR [%/y] | 7.01 | 5.163 | 4.931 | 2.405 | 9.277 | 9.514 |

**Table 2:** T = 10 years run

As stated, in the computations above, trading fees are neglected. An approximate upper bound $C$ for the maximum trading cost per trade can however easily be calculated based on the CAGR of a strategy as $C = CAGR^{\Delta t_r/250}$ where $\Delta t_r$ is the portfolio rebalancing time, i.e. the number of days between successive trades / portfolio adjustments. As the ECO strategies trade daily, we have $\Delta t_r = 1$. For the other strategies, we have only a single initial trade, as their investment fraction remains constant. Thus, a significantly lower number of trades must be carried out for these strategies, at least in our simulation setting. Nevertheless, in real applications, typically also the fraction of CK strategies is re-estimated and adjusted on a regular basis.

Table 3 gives the maximum allowed trading fee per trade $C$ for a strategy with a given CAGR to still be profitable, in units of basis points (1bp = 0.01%). The values are computed based on the CAGR values from Table 1 & 2. For $\Delta t_r = 1$, the computed value is quite low, such that typical trading fees of the order of 10bp per trade would likely eat up the gains of these strategies. However, as in practice, we typically do not rebalance the portfolio on a daily basis, we also compute the trading cost for a rebalancing period $\Delta t_r = 10$ business days. The corresponding average return per rebalancing period $\Delta t_r = 10$ business days is typically ten times larger at approximately 40 bp, allowing for trading fees of 10 or 20 bp and still generate profits. The other strategies face much lower trading fees, as they only trade once. However, in the long run case of 10 years, the ECO strategies also significantly surpass the CAGR of the other strategies. For instance, for a return of 26 bp/trade after subtraction of fees of 10bp, for the constrained ECO strategy, at the estimated CAGR and when trading every 10 business days, a return of about +91% would be achieved over 10 years, compared to a lower return of about +62% for the limited CK strategy. Further characteristics that make the ECO strategy more eligible as a trading strategy compared to CK are given in the following.

|  |  | $ECO_{[-1,2]}$ | $ECO_{[-\infty,\infty]}$ |
|---|---|---|---|
| $N$ = 2 years | $\Delta t_r = 1$ | 4 | 4 |
|  | $\Delta t_r = 10$ | 40 | 43 |
| $N$ = 10 years | $\Delta t_r = 1$ | 4 | 4 |
|  | $\Delta t_r = 10$ | 36 | 36 |

Table 3: Maximum possible trading fees per trade in basis points.

To present a yet more aggregate view of the strategy performance results presented in Table 1 & 2, we next sweep the window duration T between 250 points (representing one year of trading data) and 10,000 points (40 years) in steps of 250 points. At each window duration, again, we simulate $m = 10,000$ synthetic price series at the base parameter values $\Phi$ (11). For each of these individual simulations, we run all six portfolio strategies to obtain the corresponding investment fraction series $\{\lambda\}_{ij}$ and the wealth / portfolio evolution time series $\{W\}_{ij}$ where the subscript $i = 1, ...,6$ indicates the i'th strategy and $j = 1, ..., m$ indicates the j'th simulation. For each simulation, we find the terminal log-wealth of each strategy $W_{T,ij}$, as well as the fraction of uptime (i.e. the fraction of time in the simulation time window of length $T$ that the strategy is located above the underlying price). Additionally, we compute the mean investment fraction of a given strategy as the sample mean $\bar{\lambda}_{ij} := \text{Mean}(\{\lambda\}_{ij})$. We then calculate the median, lower and upper quartiles of the computed metrics over all $m$ simulations.

Figure 5 compactly shows the quantiles and medians plotted over the tested window durations for all metrics. Therefore, each point in the figure now represents a set of the previously seen histograms. This allows us to observe the general trend of the quantities in a compact view. The shaded, coloured areas indicate the interquartile ranges of the metrics distributions.

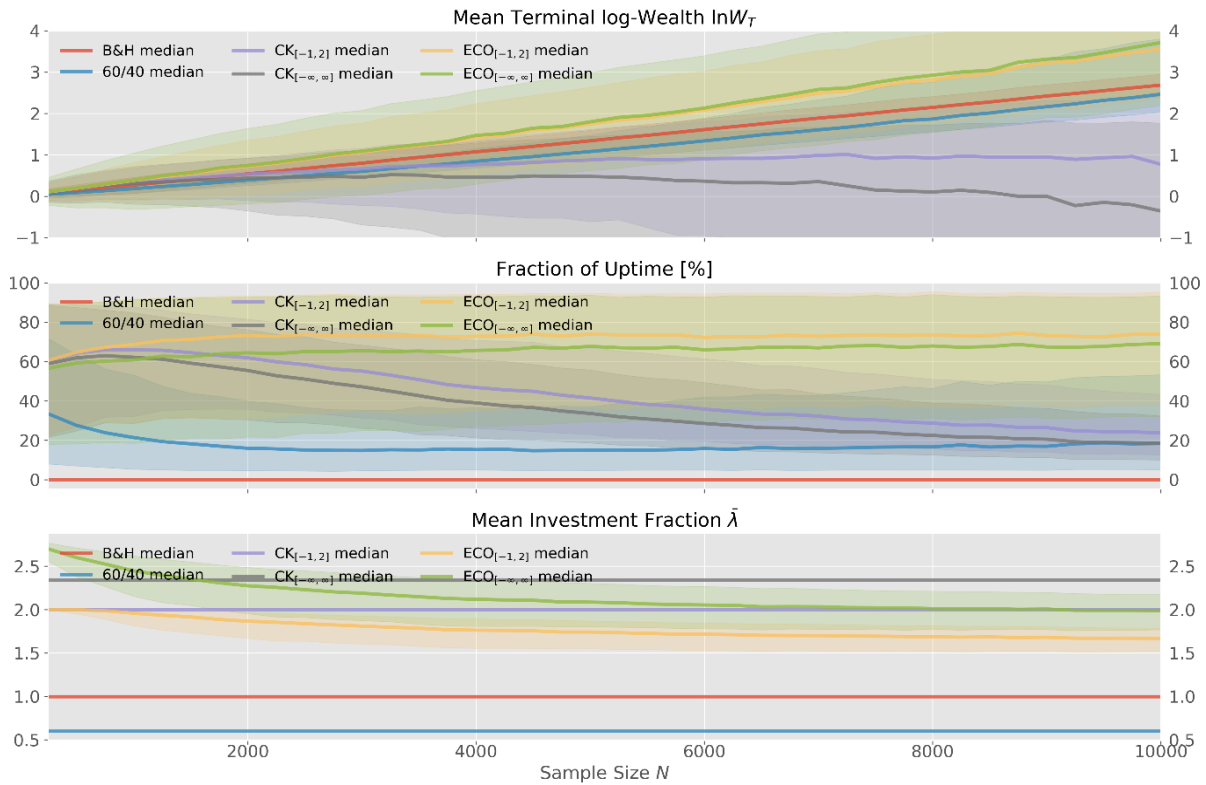

**Figure 5:** Comparison of different median strategy performance metrics for the six strategies, as simulated in a Monte-Carlo run of 10000 simulations per window duration.

In the top panel, we see clear outperformance of the ECO over the CK strategies (note the "log-unit" of the log-wealth), as also demonstrated in the histogram and table studies before. Furthermore, above a time window size of about 4000, the CK strategies clearly underperform the simulated price series which show their failure under large leverage and in case of data that contains large price jumps. This demonstrates that, even in the rare presence of jumps ($\rho = 0.01$), it already makes a striking difference in the long run whether we incorporate this risk into our strategy.

As visible in the second panel of Figure 5, the ECO portfolios manage to attain outperforming returns, as well as robust and stable fractions of uptime of about 70-80%. Thus, with ECO strategies, we do not only obtain larger terminal values at a higher probability, but they also result in portfolio values that exceed the value of the B&H strategy the major part of the time. Thus, whenever investors decide to close their portfolios, they would end up with a profit in 70-80% of the time (for the given parameters).

We pointed out earlier that investors may risk increased probability of bankruptcy when they introduce leverage into the Kelly strategy (and in return receive potentially larger payoff). As we show in Figure 6 below, for the classical Kelly strategy, an increasingly large fraction of strategies can end up in bankruptcy, as shown by the probability of default, which measures exactly this (see definition above). This is an already well-known side-effect of Kelly strategies applied to pure GBM data. Essentially, high employed leverage "into the wrong direction" or simply an unfortunate series of bad outcomes (MacLean et al., 2010c) may be the cause of total bankruptcy. On the contrary, as the figure reveals, when we invest according to the ECO strategies, the fraction of bankrupt strategies is significantly lower for all window durations (though also slowly increases).

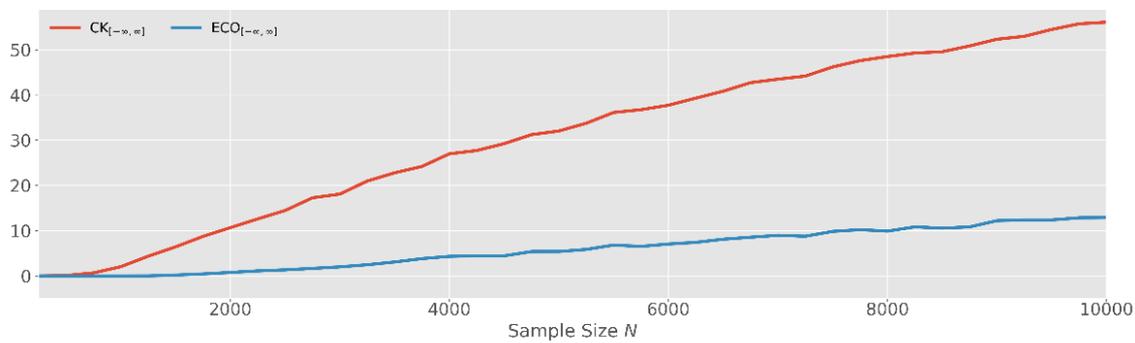

**Figure 6:** The probability of default (i.e. the fraction of strategies ending in total bankruptcy of the investor) for the unbounded CK and ECO strategies over various time window sizes. There is an increasing trend in both series, however it is much more extreme for the CK strategy.

The main conclusions of a simulation study of Ziemba (2016) for the Kelly pure GBM strategy are:

a. the great superiority of full Kelly and close to full Kelly strategies to other strategies over longer horizons with very large gains a large fraction of the time;
b. the short-term performance of Kelly and high fractional Kelly[5] strategies is very risky;
c. there is a consistent tradeoff of growth versus security as a function of the bet size;
d. no matter how favorable the investment opportunities are or how long the finite horizon is, a sequence of bad scenarios can always lead to very poor final wealth outcomes, with a loss of most of the investor's initial capital.

We demonstrate similar properties for our augmented Kelly formulation, such as long-term outperformance, maximization of the median log of wealth, as well as trading off security (bounded $\lambda$) for lower growth (MacLean et al., 2010c), however with a significantly lower probability of default and thus lower extreme downside risk. This results mainly from the important additional advantage of our strategy to incorporate crash risk in terms of the expected crash size, compared to other strategies.

---

[5] The fractional Kelly strategy means to only put a fraction of the proportion of wealth obtained by following the Kelly criterion into the risky asset, which makes the results less volatile but may not be optimal in the long run.

The observations made up to this point are based on two assumptions, (i) data generated from the EC model matches real-world data more accurately than GBM data and (ii) the true parameter values are fixed and known. If (i) is true, then one should prefer the Kelly strategy derived from our model specification over the classical Kelly strategy based on the GBM data model. However, our model involves more parameters, and thus potentially more realistically captures the behaviour of real data. We need to determine the dependence of the ECO Kelly procedure on the underlying values of the model parameters, as well as the influence of the estimation error on these parameters. Thus, in the following section, we perform a parameter sensitivity analysis, given the known parameter values, and then relax assumption (ii) by introducing estimation error.

## 5. Sensitivity Analysis

So far, the presented results were simulated based on a fixed set of model parameter values that we considered a realistic choice. In this section, we focus on the performance of the ECO portfolio as a function of differing parameter values. Thus, we regard the performance as model output. Again, we assume that the true parameters are perfectly known. This assumption will be relaxed later.

The sensitivity analysis procedure is as follows. We keep the set of base parameter values defined in (11). Our main parameters of interest are $r_D, r_N, \sigma, \rho, \overline{K}$. We scan each of them over a reasonable search range. While keeping the values of the other parameters fixed at their base values, we vary the current parameter over its associated search range. At each point in the parameter search range, we generate $m = 10{,}000$ synthetic price paths of duration $T = 2{,}500$ time steps (i.e. 10 years) with the current parameter values. We then run the ECO portfolio strategies with constrained ($\lambda \in [-1,2]$) and unconstrained leverage and, as above, compute the corresponding distributions of the terminal log-wealth, the fraction of uptime and the mean investment fraction over all simulations. In Figure 7 - Figure 10, we depict the medians and interquartile ranges of the resulting distributions, however in this case as a function of the sweep parameter value instead of the time window size (which is fixed). In addition to the above-mentioned metrics, we also compute the probability of default (fraction of bankrupt strategies) for each parameter value. The other performance metrics are computed conditional on all non-bankrupt strategies. Note that the results for $r_D$ and $r_N$ are summarized in one Figure. They will be identical, as $r_N = r_D$ throughout all simulations.

Evaluating the results parameter by parameter yields several insights, which we now describe.

**Volatility σ (Figure 7):** As volatility increases, we still obtain positive performance. This is also indicated by the rising fraction of uptime. On average, the investment fraction decreases at higher volatility. The surprising fact is that the ECO strategies manage to outperform, which is explained by its knowledge of the mispricing component. Knowledge of the mispricing (i.e. the bubble size) thus allows investment decisions leading to positive performance, independent of the drift of the data and only as a function of the realised volatility. Furthermore, as expected, we obtain that the probability of default increases with volatility, however it seems to flatten at a level of around 20-25% for the parameters used in our study.

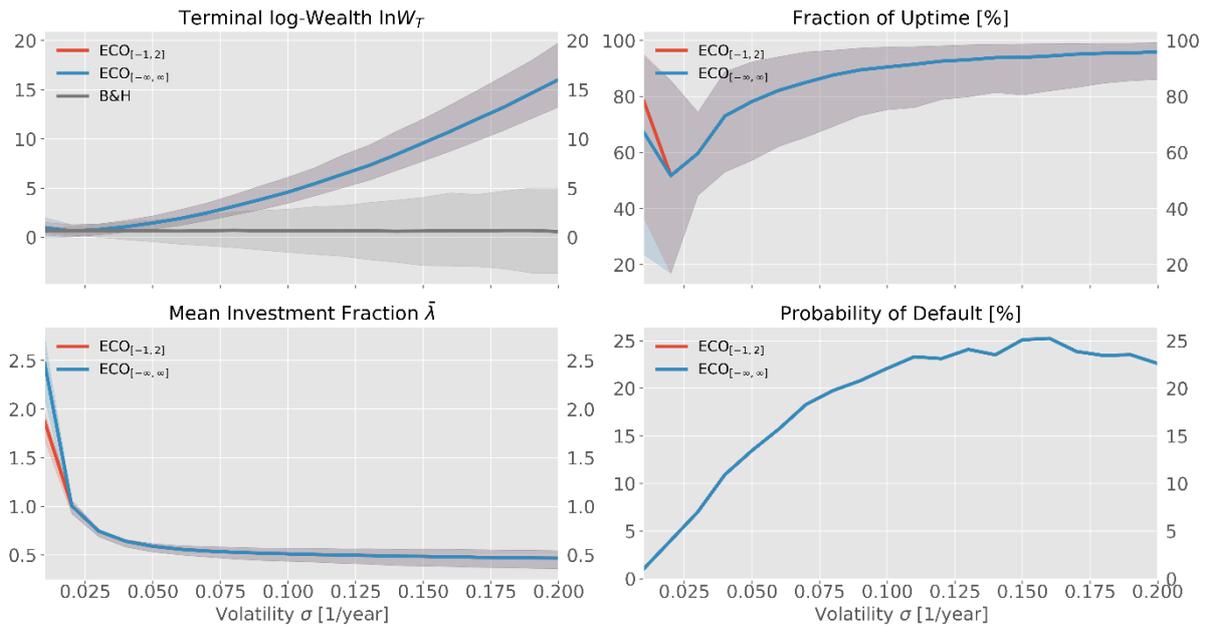

**Figure 7:** Sensitivity analysis results for the parameter $\sigma$.

**Price Process Rates $r_D$, $r_N$ (Figure 8):** The sensitivity analysis for the growth rates clearly demonstrates that the ECO strategy performs well at stronger trends in the underlying price series. However, this only holds for the constrained strategy. In the unconstrained case, the applied leverage becomes too large: on the one hand, this yields to strong performance of winning strategies, however on the other hand, the probability of default increases extremely. In case of the constrained strategy, we obtain lower growth, however at a larger fraction of uptime and significantly lower probability of default. This result is no surprise and demonstrates the importance of employing a limited leverage approach.

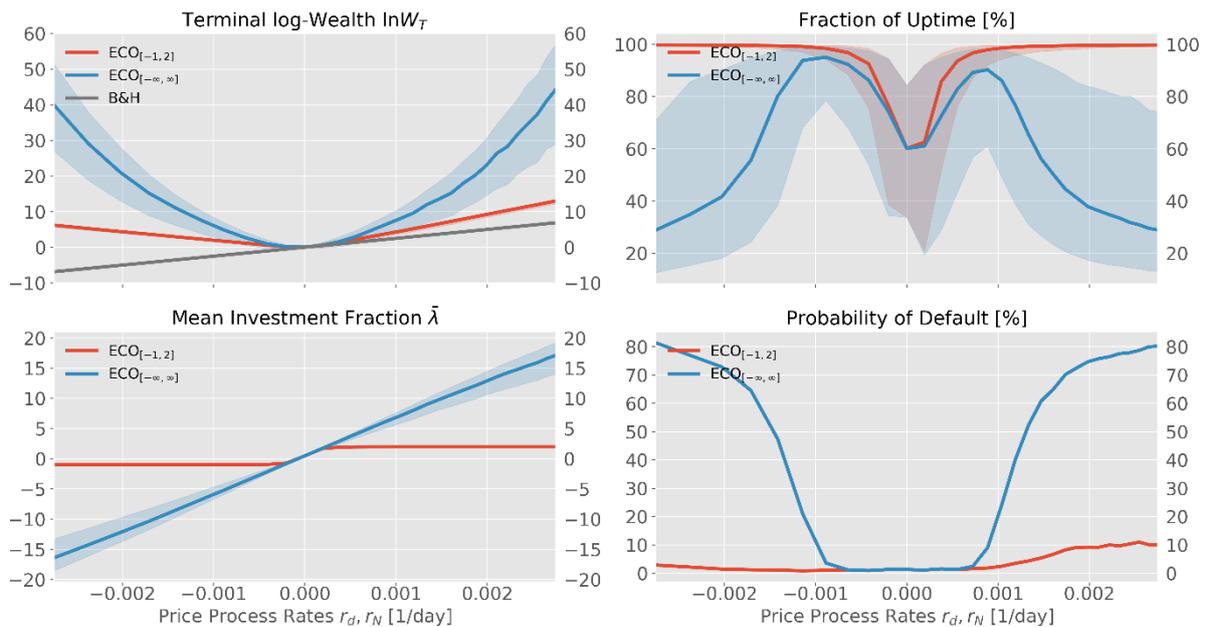

**Figure 8:** Sensitivity analysis results for the parameters $r_D$ and $r_N$.

**Jump Probability ρ (Figure 9):** The probability of encountering a jump has a weaker impact on the overall performance. Nevertheless, there are slight increases in performance, but also in the probability of default, at higher values of ρ. Furthermore, as the probability increases, on average the investment fraction is corrected to lower values, as desired.

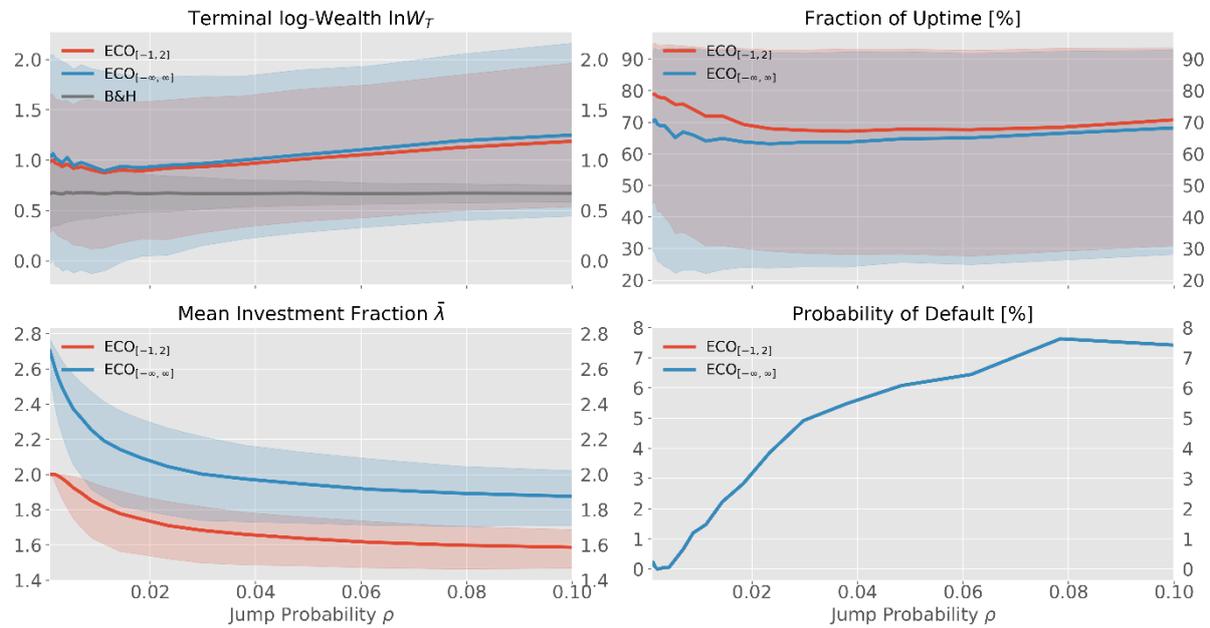

**Figure 9:** Sensitivity analysis results for the parameter $\rho$.

**Average Corrective Jump Size $\overline{K}$ (Figure 10):** Lastly, the sensitivity analysis of the jump size reveals decreasing performance at increasing jump sizes. The investment fraction is above one for all tested values of $\overline{K}$. Thus, jumps are intensified, which leads to an increasing fraction of strategies ending in bankruptcy. In real applications, we damp this effect by fixing $\lambda \in [0,1]$ or $[-1,2]$.

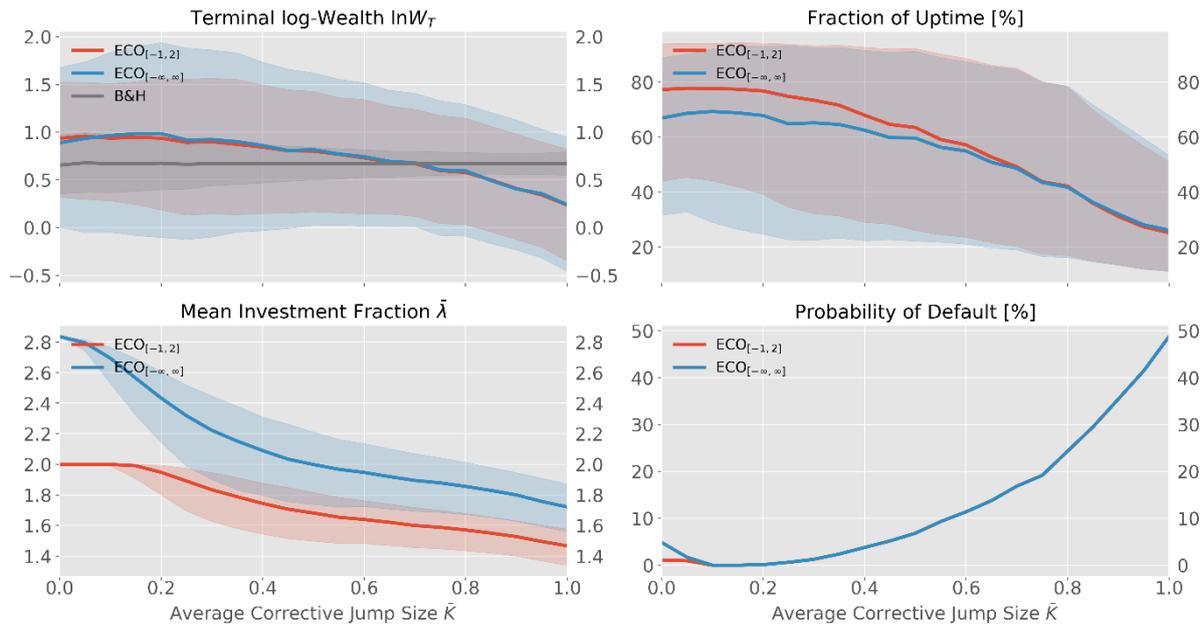

**Figure 10:** Sensitivity analysis results for the parameter $\bar{K}$.

## 6. Estimation Error Analysis

Until now, we have presented the performance of the ECO portfolio and other standard strategies assuming the true dynamics of the underlying data are known. Now, we move to a more realistic setting by imposing additional Gaussian estimation error to the true parameter estimates. We thus generate again synthetic time series at the true parameter values, however we run the ECO portfolio on this data given faulty estimates of the true parameters. By doing so, we determine the sensitivity of performance in dependence of the imposed level of error. We can therefore directly draw conclusions about the estimation error, as the model error under the synthetic data approach is zero.

The robustness test is carried out for each model parameter separately in the following way. In order to determine the dependence of the strategy performance on the magnitude of the error, we sweep the Gaussian error standard deviation $\sigma_e$ over a logarithmically spaced range from $10^{-3}$ to $10^2$. For each tested value of the standard deviation, we generate $m = 10,000$ price series at the true base parameter values. Then, we run the ECO over these price series, given the base parameter values (11), plus an incorrect estimate of the currently analysed parameter:

$$\phi_e = (1 + \varepsilon_i)\phi_{true} \qquad (14)$$

where $\varepsilon_i \sim N(0, \sigma_e^2)$, $i$ is the number of the current underlying price series, and $\phi_{true}$ is the base value of the currently analysed parameter[6]. Thus, in each simulation $i$, we run the portfolio with a new, different error estimate of the true parameter. On average the faulty estimates are distributed normally around the true parameter value. This is reasonable when we assume that our estimation procedures for the respective parameters are unbiased / consistent.

As before, the metrics that we choose to evaluate portfolio performance are the terminal log-wealth, the fraction of uptime, the mean investment fraction, as well as the probability of default. The results of the error robustness test for each model parameter are separately depicted throughout Figure 11 - Figure 15.

**Volatility σ (Figure 11):** There seems to be a stable performance up to an order of the error standard deviation between $10^{-1}$ to $10^0$. Above this level, a drastic decline in performance occurs, which shows the level of error above which the estimation procedure 'breaks down'. Firstly, the terminal log-wealth and the fraction of uptime strongly decay. Furthermore, the probability of default strongly rises for the unbounded strategy while it remains at a stable level for the constrained leverage case. The distribution of the mean investment fraction strongly widens for the unbounded case, indicating higher uncertainty in the choice of the investment fraction. Again, the results underline the importance of utilizing a constrained leverage approach, as otherwise, we obtain an extreme number of bankrupt strategies.

---

[6] We constrain the error estimates of $\sigma, \rho$ and $\overline{K}$ to be $\sigma \geq 0, \rho \in [0,1]$ and $\overline{K} \geq 0$, as these are natural lower and upper bounds for these parameters, either by their 'financial meaning' or by construction of the model. Thus, whenever an error estimate outside these limits occurs, we set it to the nearest boundary value.

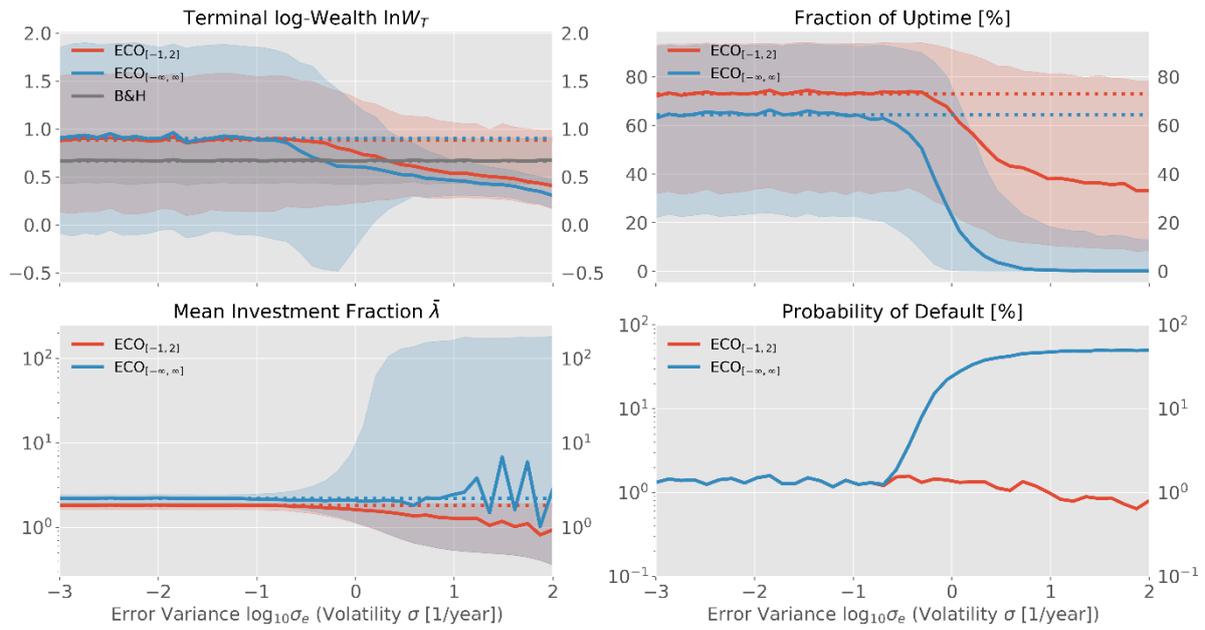

**Figure 11:** Error analysis results for the parameter $\sigma$.

**Discount Price Process Rate $r_D$ (Figure 12):** Similar but more extreme behaviour as for the volatility is obtained. Especially the probability of default strongly increases for the unconstrained leverage case. Again, the performance seems to be approximately stable up to an order of error standard deviation of $10^{-1}$ to $10^0$.

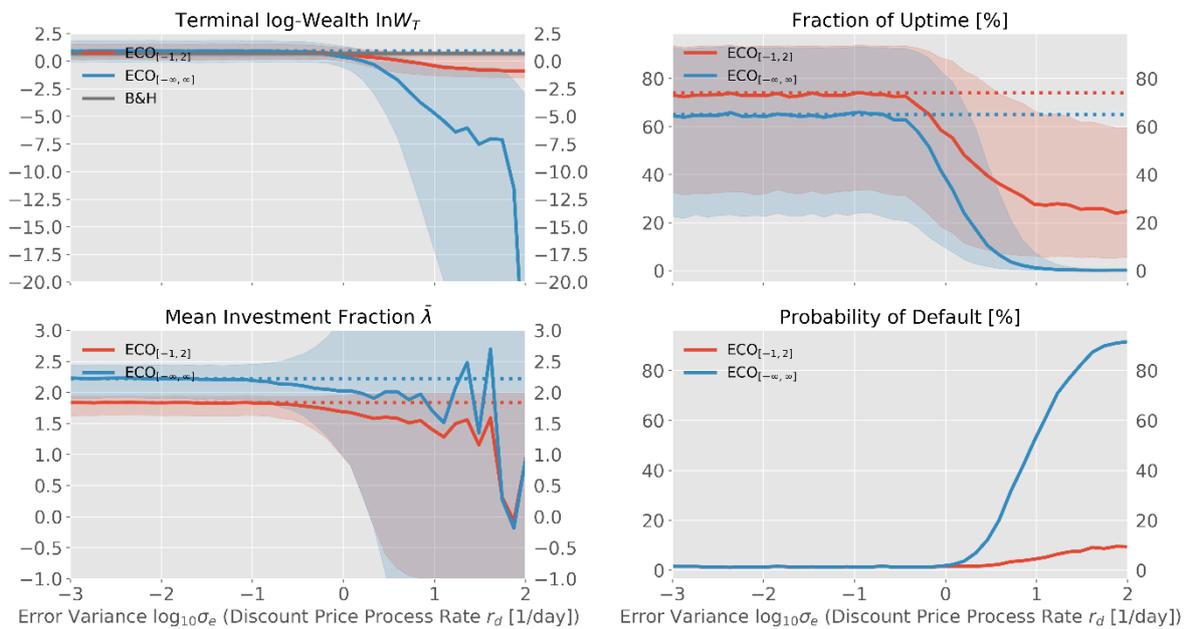

**Figure 12:** Error analysis results for the parameter $r_D$.

**Normal Price Process Rate $r_N$ (Figure 13Figure *12*):** Similar but less extreme behaviour as for the discount rate is obtained. Again, the performance seems to be approximately stable up to an order of error standard deviation of $10^{-1}$ to $10^0$.

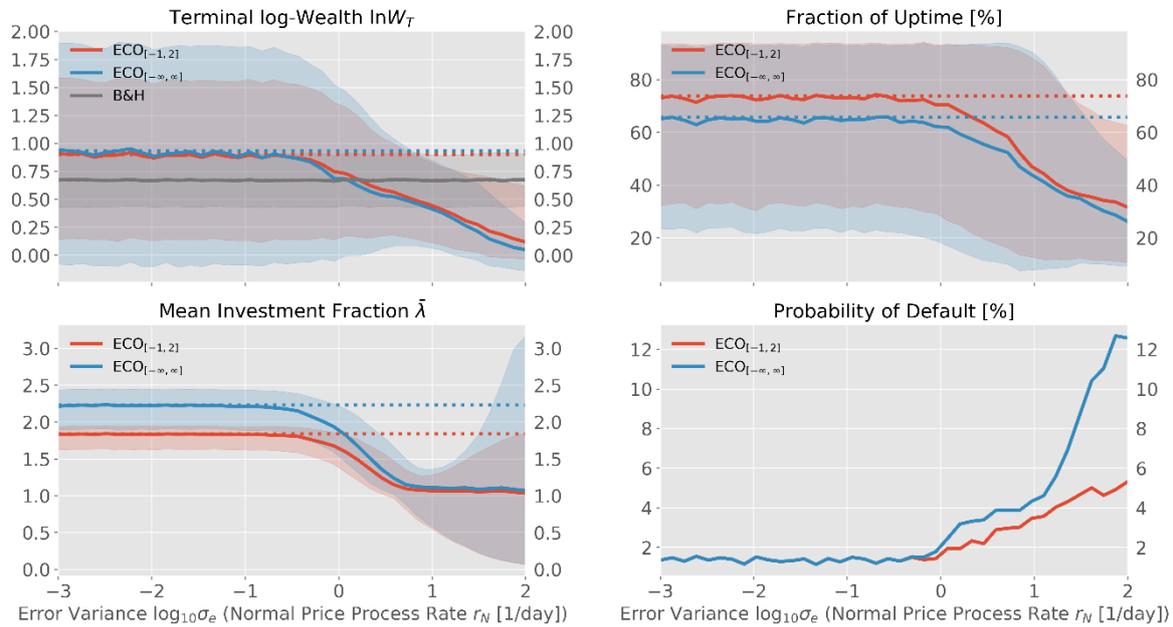

Figure 13: Error analysis results for the parameters $r_N$.

**Jump Probability ρ (Figure 14) and Average Corrective Jump Size $\overline{K}$ (Figure 15):** The dependence of the performance on the error level shows very similar behaviour for these parameters. The break points in performance are at the same order as for the other parameters, however the intensity of the performance decay is much lower.

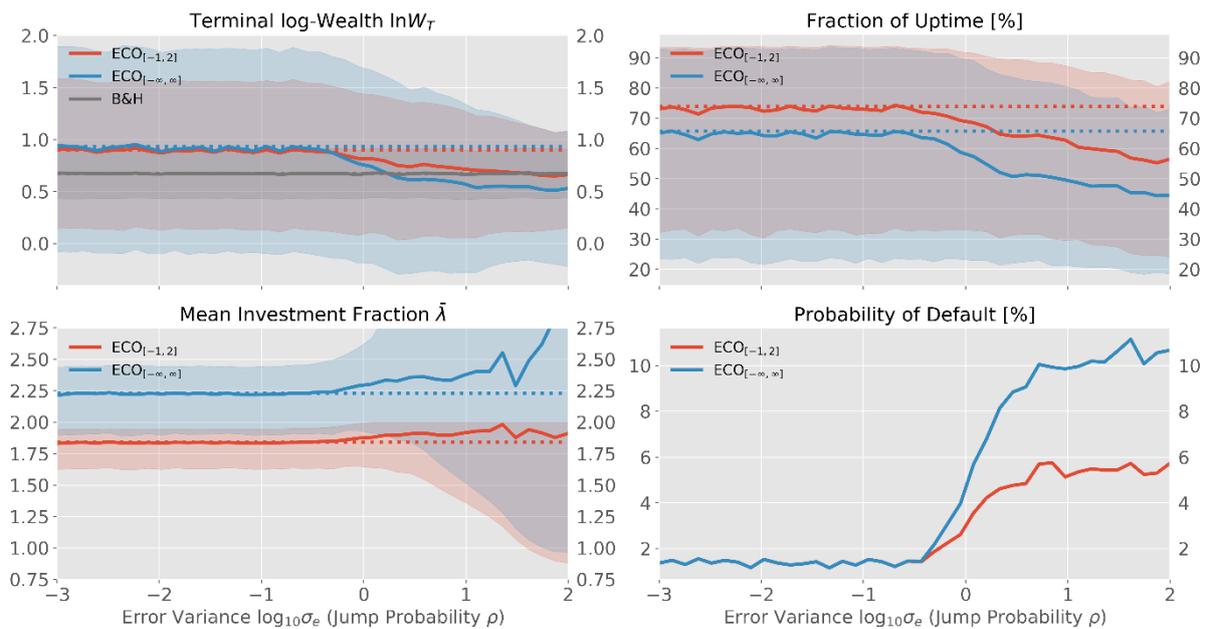

Figure 14: Error analysis results for the parameter $\rho$.

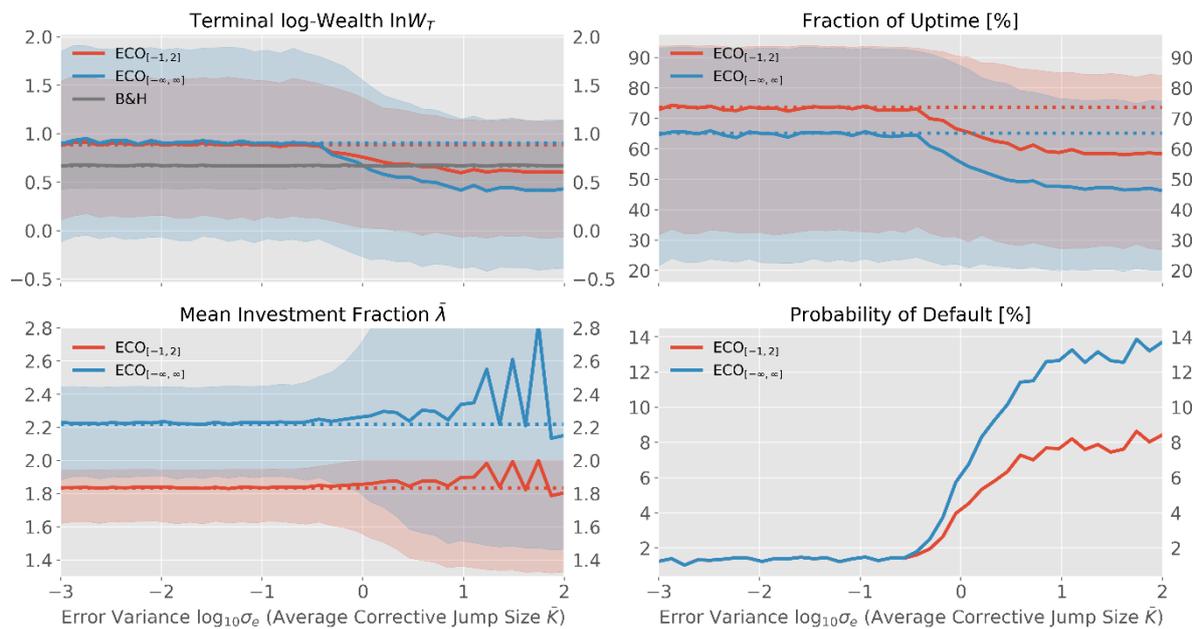

**Figure 15:** Error analysis results for the parameter $\overline{K}$.

Overall, we identify stable positive performance of the ECO portfolio up to the order of imposed estimation error standard deviation of $10^{-1}$ to $10^0$, i.e. 10%-100% error. This is a quite high level of estimation error. Thus, there is a quite good chance at deriving optimal investment decisions, even if the estimates of the parameters are faulty. Furthermore, we find that model performance is most sensitive to the correct estimation of the short-term rate $r_D$. The expected return is not only the most important parameter, but also typically the most difficult one to estimate Merton (1980). However, our error robustness analysis reveals that some variability is allowed in estimating the expected returns in ECO. The sensitivity to expected returns in trading strategies is also examined in Chopra and Ziemba (1993).

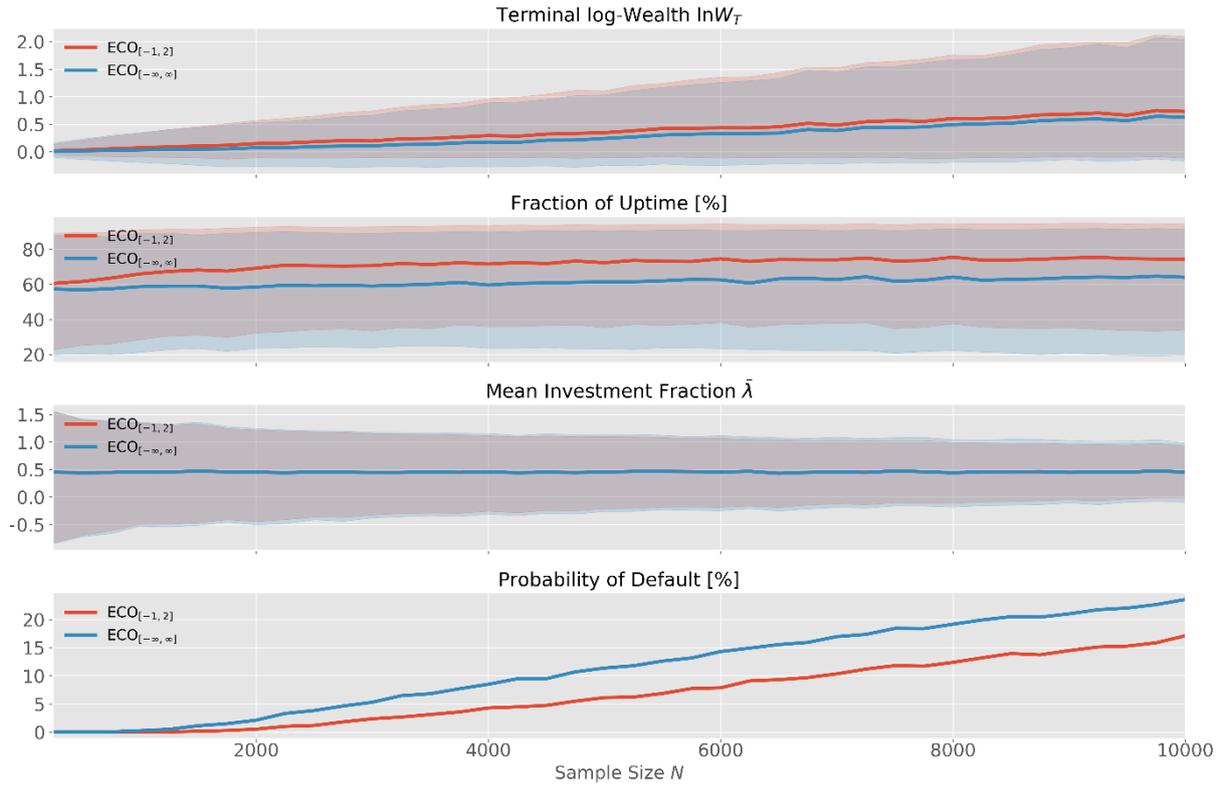

**Figure 16:** Portfolio performance under correctly estimated signs of the process rates.

**Importance of correctly estimating the sign of the drift trend**: We now demonstrate the importance of correct estimation of the sign of the drift trend in the underlying data. As in the tests before, our four performance metrics are compared over various window durations for m = 10,000 simulations per window duration. For each simulation instance, we draw a random sample $r_S$ for the process rates from a uniform distribution with limits [-10%,10%] annual growth. The true process rates are assigned the sample value, i.e. $r_D = r_N = r_S$. The other model parameters are kept fixed at the base parameter values (11). We then generate a price path of length $T$ given those parameter values. Next, as our error estimates for the process rates, we draw a new rate $r_e \sim U(0, 2|r_S|)$, such that the error estimate may be maximally 100% off the true value. Then, in order to test the dependence on the estimation sign, we compute the ECO bounded and unbounded portfolio performances given the true parameters as well as the correctly signed error estimate of the rate, i.e. $\text{sgn}(r_S) \cdot r_e$, or the incorrectly signed estimate, i.e. $-\text{sgn}(r_S) \cdot r_e$. The results are plotted in Figure 16 and Figure 17. A clear difference in performance is evident. This is especially visible in the fraction of uptime. This underscores the importance of the correct estimation of the direction of the underlying trend in the data. When this is achieved, as the procedure shows, even at errors up to 100%, the estimation may still yield excellent outcomes in terms of performance. When the sign of the process rates is wrongly estimated, we obtain bad performance, as the sign of the investment fraction is also strongly influenced by this.

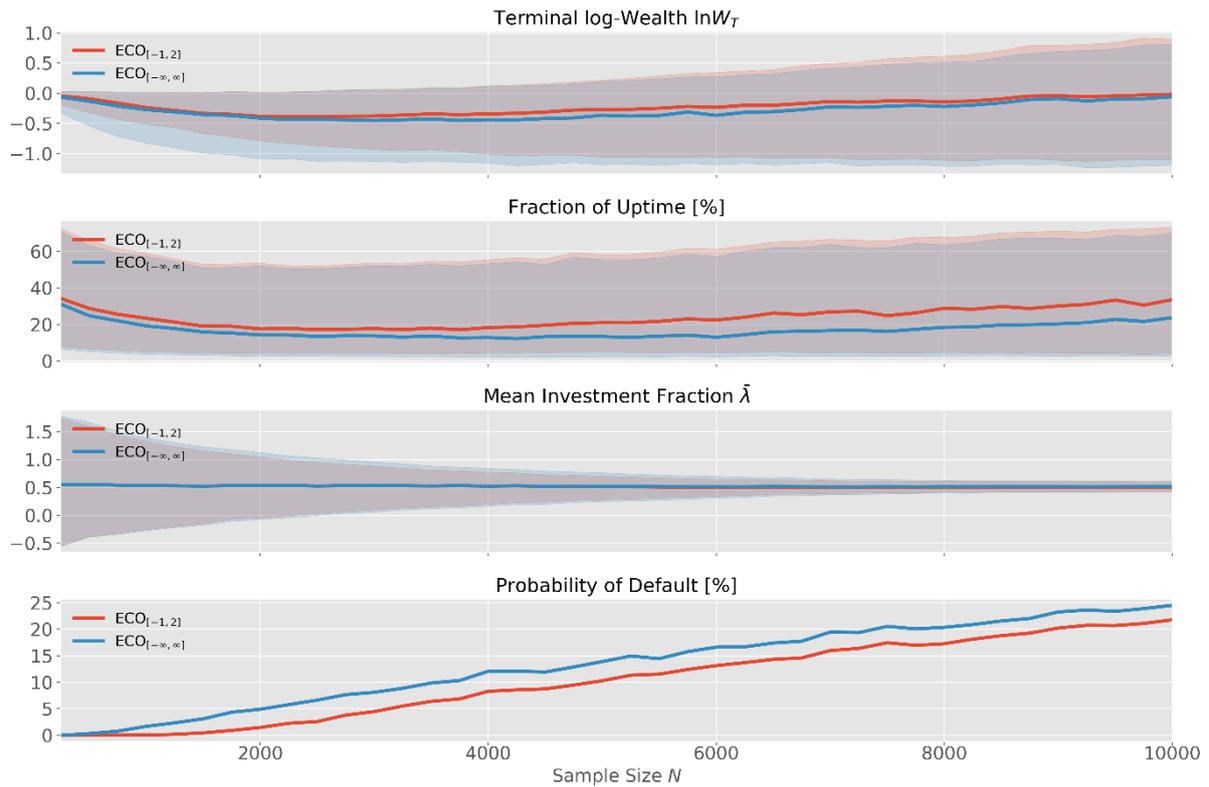

**Figure 17:** Portfolio performance under incorrectly estimated signs of the process rates.

This result, as well as the other results stated in this section, essentially show us that, although parameter estimation has a large impact on the performance, we can achieve reasonably well performing portfolios also under faulty parameter estimations.

In reality, faulty parameter estimates are very likely to occur and the employed estimation procedures are often the bottleneck in performance of complex models with many parameters; especially with financial data, we do not have the convenient option to arbitrarily often "repeat the experiment" until we accumulate sufficient data such that we can average the correct parameter values. Instead, we only have a single instance of a random process, the return series (and possibly a few correlated time series) given and we need to extract as much information about the true parameters from it as possible. We also need to pay attention to not overfit to the particular given instance of the random process, as it may be completely different from the global behaviour of the underlying true random process and we may be prone to extracting spurious relationships from data. These may work in hindsight, in order to explain in-sample performance, but are not robust with respect to the future. As we show in this work, we however find that, if we can do the job "fairly well", i.e. within a realistic range of estimation error, we can in fact produce rewarding strategies. As we see from the dependence of the performance on the underlying value of the discount price process rate $r_D$, this is especially the case when there is a strong trend in the data. Thus, as commonly said, the trend is your friend, and application of the model to such assets should likely result in winning portfolio strategies if we can estimate the corrections.

## 7. Conclusion

In this paper, we have examined the performance of the Efficient Crashes method without price acceleration based on synthetic data. We have shown that the method performs well and as expected on synthetic data. Through the analyses, we have gained more insights into the model parameter and estimation quality impact on the portfolio performance, as well as the performance compared to other standard strategies. Most importantly, the method is quite robust in the parameter estimation error, making it more tractable on real data. A robust model specification such as ours, which exhibits a "buffer" for the allowed estimation error, is quite desirable, as it will produce rewarding strategies for a broad spectrum of estimation settings.

Furthermore, we have shown that the augmented Kelly strategy that we derive based on our model exhibits behaviour like classical Kelly on GBM. Specifically, the method

- outperforms buy & hold, 60/40 and classical Kelly strategies applied to synthetic data with jumps;
- is robust in the estimation of parameter values;
- improves its performance in the long term;
- beats the buy & hold strategy about 62% of the time over a two-year period and 65% over a ten-year period (with limits on the $\lambda$ values);
- performs like the classical fractional Kelly strategy in controlling volatility (with limits on the $\lambda$ values);
- on average beats all the other methods in terms of CAGR on a two-year and ten-year basis; and
- shows an improved Sharpe Ratio over classical Kelly on a two-year and ten-year basis.

Our augmented Kelly strategy is aware of the presence of two sources of risk, namely (i) the "normal" volatility occurring in asset time series and (ii) the risk of a crash. With the additional, useful assumption of "efficient crashes", one can split and incorporate the risk of a crash in two separate ways; its time of occurrence as well as its size. Here, we examine only the latter of the two and show that exploiting this information is already sufficient to attain outperformance compared to strategies that do not consider crash risk. The presented method is an improvement over classical Kelly, when applied to price paths with GBM plus jumps relative to a normal price (which we deem a more realistic model of real-world financial time series behaviour).

We did not explore the impact of a dynamic model of the probability, as we consider this a separate building block of the model. We furthermore did not look at the model performance when the underlying data is generated with dynamic parameter values, in particular for the expected return and volatility, yet. This would require the specification of additional models or procedures to generate dynamically changing values for these parameters, which exceeds the targets of this study. In future work, we plan to explore the model performance on a larger variety of historical bubble (and non-bubble) datasets in ex-ante simulations, in order to complement further the first successful study results on real-world datasets from the initial paper Kreuser and Sornette (2018).

## 8. Appendix

**Appendix A: Approximate Solution for the Optimal Kelly Investment Fraction**

For simplicity, we use $\rho$ in place of $\bar{\rho}$ or $\rho_t$ in the following. The solution of the optimization problem to obtain the asset allocation $\lambda^*$ reads:

$$\lambda^* = \frac{D\left(1 + \frac{\sigma^2}{2}\right) - 1}{(1-\rho)(A-1)^2 + \rho(B-1)^2 + H_2 + H_3}$$

$$\approx \frac{r_D - r_f + \sigma^2/2}{\sigma^2 + (1-\rho)(\bar{r} - r_f)^2 + \rho(\bar{K}\ln q_t + r_D - r_f)^2}$$

with

$A = e^{\bar{r} - r_f}$

$B = e^{\bar{K}\ln q_t + r_D - r_f}$

$D = (1-\rho)A + \rho B$

$H_2 = \left(2\left((1-\rho)A^2 + \rho B^2\right) - D\right)\sigma^2$

$H_3 = \left((1-\rho)A^2 + \rho B^2\right)\frac{3}{4}\sigma^2$

## Appendix B: Price Trajectory Examples for various Parameter Settings

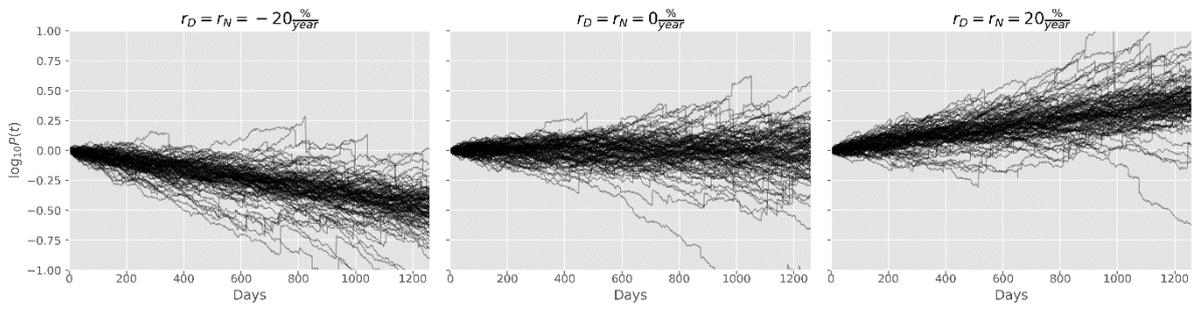

**Figure 18:** Dependence of the synthetic data trajectories on the parameters $r_D$ and $r_N$.

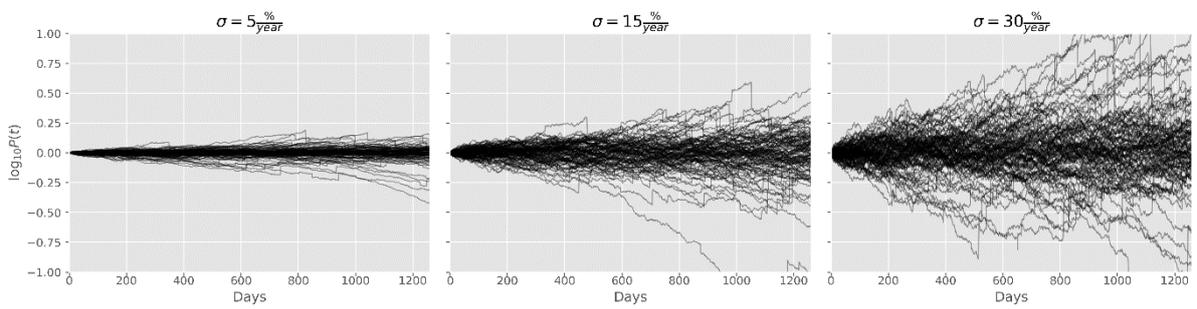

**Figure 19:** Dependence of the synthetic data trajectories on the parameter $\sigma$.

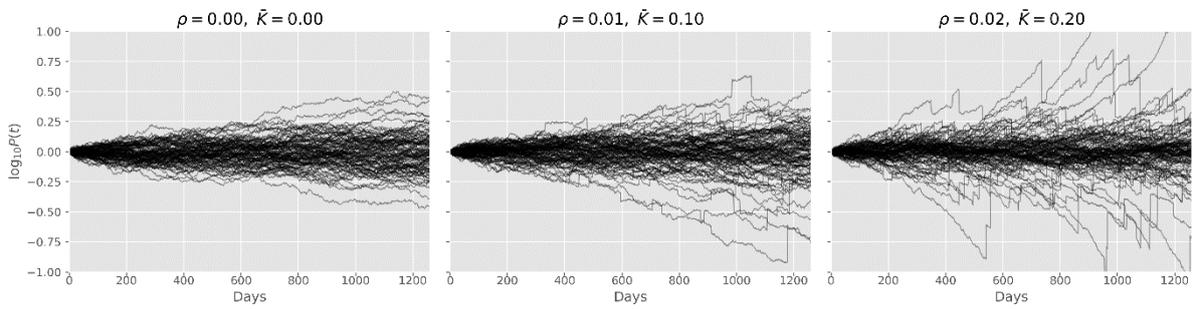

**Figure 20:** Dependence of the synthetic data trajectories on the parameters $\rho$ and $\overline{K}$.